\documentclass[reprint, amsmath, onecolumn, pra,superscriptaddress]{revtex4-2}

\usepackage{graphicx}
\usepackage{epsfig}
\usepackage{dcolumn}
\usepackage{bm}
\usepackage{blkarray} 
\usepackage[nameinlink, capitalise]{cleveref} 

\newcommand{\cmm}{\ensuremath{\mathrm{cm}^{-1}}}
\newcommand{\Xstate}{$X^1\Sigma^+$}

\newcommand{\astate}{$a^3\Sigma^+$}

\newcommand{\Estate}{$E$(4)$^1\Sigma^+$}
\newcommand{\comment}[1]{}
\graphicspath{ {./images/} }

\newcommand{\tj}[6]{ \begin{pmatrix}
    #1 & #2 & #3 \\
    #4 & #5 & #6
  \end{pmatrix}}

\begin{document}

\title{Intensities of KCs $E(4)^1\Sigma^+\to (a^3\Sigma^+,X^1\Sigma^+)$ band system up to dissociation threshold: an interplay between spin-orbit, hyperfine and rovibronic coupling effects}

\author{I. Klincare}
\author{M. Tamanis}
\author{R. Ferber}
\affiliation{Laser Center, Faculty of Physics, Mathematics and Optometry, University of Latvia, 19 Rainis blvd, Riga LV-1586, Latvia}

\author {E. A. Pazyuk}
\author{A. V. Stolyarov}
\affiliation{Department of Chemistry, Lomonosov Moscow State University, 119991, Moscow, Leninskie gory 1/3, Russia}

\author{I. Havalyova}
\author{A. Pashov}
\affiliation{Faculty of Physics, Sofia University, 5 James Bourchier Boulevard, 1164 Sofia, Bulgaria}

\date{\today}

\begin{abstract}
The relative intensity distribution in the rotationally resolved laser-induced fluorescence spectra belonging to the $E(4)^1\Sigma^+\to (a^3\Sigma^+_1,X^1\Sigma^+)$ band systems of the KCs molecule was analyzed. The experimental intensities in doublet $P$,$R$ progressions assigned to spin-allowed $E\to X$ and spin-forbidden $E\to a$ transitions up to their common ground dissociation limit were described in the framework of a coupled-channels (CC) deperturbation model applied for the interacting \Xstate\ and \astate\ states. The CC intensity simulation was based solely on fixed electronic structure parameters as functions of the internuclear distance $R$, namely: accurate empirical potential energy curves for all three states, \emph{ab initio} estimates for matrix elements $A(R)$ of the hyperfine structure (HFS) , and transition dipole moments $d_{EX}(R)$ and $d_{Ea}(R)$. A comparison between the measured intensities and their theoretical counterparts demonstrates a strong competition between different intramolecular interactions. A weak spin-orbit coupling of the upper \Estate\ state with the remote $^3\Pi$ states is responsible for appearance of the $E\to a$ vibrational bands for the intermediate $v_a$-values. In turn, the HFS coupling between \Xstate\ and \astate\ states leads to peculiarities in $E\to (X,a)$ intensities, which are pronounced for high $v_{X/a}$-values in the vicinity of K(4$^2S$)+Cs(6$^2S$) dissociation threshold. Both adiabatic ro-vibrational and non-adiabatic electronic-rotational interactions explain the abrupt deviation of some observed $P/R$ intensity ratios from the expected H\"{o}nl-London factors.
\end{abstract}

\maketitle

\section{Introduction}

Polar bialkali molecules containing a heavier Rb or Cs atom attract great attention from cold-molecules community. One of the reasons is that when ultra-cold species of such molecules are oriented by electric field they exhibit anisotropic long-range interaction properties, which is important for quantum simulation and processing of quantum information, see \cite{Carr:2009, Ulmanis:2012, Quemener:2012, Ospelkaus:2008}.  In spectroscopic studies of such molecules a particular emphasis has been given to a proper understanding of various kinds of adiabatic and non-adiabatic interaction processes both in the excited and ground electronic states as they may have a critical impact on the probabilities of the optical transitions. Accurate information on rovibronic transition probabilities corresponding to weakly bound levels located near the ground states dissociation threshold are  required to optimize laser synthesis of the ultracold molecular ensemble by means of Stimulated Raman Adiabatic Passage (STIRAP) method \cite{Ospelkaus:2008}. Exploring such optical cycles allows for  coherent transfer of weakly-bound ultracold molecules formed by photo-association or magneto-association in the weakly bound ground triplet \astate\ or singlet \Xstate\ state levels to a deeply bound ground state level, preferably the absolute ground state \Xstate\ ($v_X$ = 0, $J_X$ = 0).

Important sources of reliable information on different kind of coupling effects are experimental studies of intensities of transitions, which involve the respective states. Although this piece of information is as valuable as the line frequencies, it usually requires much more experimental efforts and advanced computational methods. On the experimental site, one needs to account for the spectral response of the setup, background, overlapping with neighboring lines etc., therefore, the uncertainty of the experimental intensity distributions rarely is below few $\%$.

It should be reminded that the often taking place breakdown of the traditional adiabatic approximation can lead to dramatic changes in the corresponding transition probabilities \cite{Field}, and to evaluate this effect one needs non-adiabatic wave functions of the coupled states. In Ref.~\cite{Field} it is shown that for example a relatively weak spin-orbit (SO) interaction between distant singlet and triplet states can be responsible for non-vanishing singlet-triplet optical transitions. At the same time a strong local SO coupling can even change the nodal structure of vibrational wave functions of the affected states. Similarly, the centrifugal distortion systematically shifts the rovibrational energy and a nodal structure of the corresponding rovibrational wave functions of an isolated (adiabatic) electronic state, while local non-adiabatic coupling of $\Omega$ = 0 and $\Omega$ = $\pm 1$ states may cause interference effects between so-called parallel ($\Delta\Omega$ = 0) and perpendicular ($\Delta\Omega$ = $\pm 1$) rovibronic transitions \cite{Field}. Eventually, both effects will result in a non-regular (non-H\"{o}nl-London) redistribution of the intensities between $P$ and $R$ branches of the same band.

In the current work we focus on the KCs molecule, which is under active studies aimed to obtain it in ultra-cold conditions, see \cite{Patel:2014, Borsalino:2016, Groebner:2017}, because of its sufficiently large electric dipole moment of 1.92 D and stability with respect to molecular collisions \cite{Zukowski:2010}. In particular, the interspecies Feshbach resonances in ultra-cold K-Cs pairs have been experimentally observed and modeled~\cite{Groebner:2017}. In the present paper we concentrate our efforts on analysis of adiabatic and non-adiabatic effects, which determine the intensities of optical transitions from the excited \Estate\ state to the ground \Xstate\ and \astate\ states.

An empirical potential energy curve of the shelf-like \Estate\ state (see Fig.1a), which converges to the  K(4$^2$S) + Cs(5$^2$D) atomic limit was determined and presented in an analytical form in \cite{Busevica:2011} and in a point-wise form in \cite{Szczepkowski:2012}. As suggested in \cite{Klincare:2012}, this state is prospective as intermediate state to transfer ultra-cold KCs species from weakly-bound rovibronic levels of the coupled ($a,X$) states to the ``absolute'' ground state \Xstate\ ($v_X$ = 0, $J_X$ = 0) in one STIRAP process $(a,X) \rightarrow E \rightarrow X(0,0)$. The required  STIRAP intensity simulations were based on the relevant  \emph{ab initio} spin-allowed $E -X$ and spin-forbidden $E - a$ transition dipole moments~\cite{Klincare:2012}. However, in Ref.~\cite{Klincare:2012} the hyperfine (HF) interaction between the close-lying vibrational levels of \astate\ and \Xstate\ states was disregarded at modeling the experimental intensity distributions of \Estate\ $\to (a,X)$ Laser-Induced Fluorescence (LIF) spectra. The effects due to this HF interaction were observed in \cite{Ferber:2009, Ferber:2013} since the shelf-like structure of the upper state allows for $E \to (a,X)$ LIF progressions ending on the levels close to the common dissociation limit of separated K and Cs ground state atoms. Ref.~\cite{Ferber:2013} presents a deeper insight into the mixing between the ground state $v_X/v_a$ vibrational levels mainly due to Fermi contact interaction coming from the Cs atom's large nuclear magnetic moment. As recommended in Ref.~\cite{Ferber:2013}, a more comprehensive deperturbation treatment of a nodal structure of the $X-a$ non-adiabatic rovibrational wave functions is necessary to cover all phenomena that determine \Estate\ $\to (a,X)$ intensities. It should be also mentioned in this connection that the observed unusual intensity ratios between $P$ and $R$ lines of the same doublet have not been analyzed in previous studies \cite{Klincare:2012, Ferber:2013}, in which the calculated values were averaged over the $P$ and $R$ components of a doublet.

During previous studies \cite{Busevica:2011, Klincare:2012, Ferber:2013} a large number of $E(v^\prime, J^\prime) \to (a,X)$ Fourier-transform (FT) LIF spectra were obtained in the University of Latvia Laser Center. The recorded spectra combine high spectral resolution with high detection sensitivity, which allowed for determining LIF intensity distributions including rather weak lines. A rich amount of not yet explored information on intramolecular interactions in ground and excited states contained in these spectra motivated us to perform a detailed analysis of the intensity distribution that would allow for testing of the applied deperturbation model. In a recent paper \cite{Krumins:2022} we presented a simplified $2\times2$ coupled channels (CC) model for the HF structure (HFS) of the interacting \Xstate\ and \astate\ states, which managed to accurately reproduce all existing experimental frequencies, including a weak, previously unobserved dependence of the HFS on $v_a$. It was explained by accounting for the $R$-dependence of HFS matrix elements reported in \cite{Oleynichenko:2020}. Accounting for these considerations, the present paper is aimed to use the available experimental data and new theoretical calculations to describe the $E(v^\prime, J^\prime) \to (a,X)$ LIF intensity distribution.

The present publication is organized as follows. In Section~\ref{Sec:Exp} we provide a brief summary of the experimental procedure followed by analysis of the obtained spectra. Then, in Section \ref{theory:sec} a summary of the theoretical background behind the simulation of the relative line intensities is presented, whereas the relevant details may by found in the three Appendixes. In Section \ref{res:sec} we present a comparison between experimental and calculated line intensities.

\section{Experimental details and spectral analysis}
\label{Sec:Exp}

The experiments aimed to detect the  \Estate$\rightarrow$(\astate,\Xstate) high-resolution  LIF spectra have been described in our earlier papers \cite{Busevica:2011, Klincare:2012, Ferber:2013}. In short, the KCs molecules have been produced in a linear heat-pipe. The spectra were recorded at temperatures of the heat-pipe of about 290$^\circ$ C. The admixture of about 1 mbar of Ar as buffer gas was necessary to prevent vapor condensation on the optical windows. A single mode ring dye laser (Coherent 699-21) with Rhodamine 6G dye was used to excite \Estate($v^\prime$, $J^\prime$) $\leftarrow$ \Xstate\ ($v_X$, $J_X$) transitions. The LIF was dispersed by a Fourier-transform spectrometer IFS-125HR (Bruker) with a typical instrumental resolution of 0.03 \cmm, but sometimes the resolution was increased up to 0.02 \cmm. The spectra have been limited by an optical band pass filter BP 716 to within 13 700 to 14 500 \cmm\ and detected by a photomultiplier (Hamamatsu R928). Laser excitation frequencies have been chosen within the 17 809 - 17 816 \cmm\ range and actively stabilized by a wave-meter (HighFinesse WS6). These frequencies were found in \cite{Ferber:2013} to excite strong transitions to the \Estate\ state rovibronic levels ($v^\prime = 44$ and $45$) from the low vibrational levels of the \Xstate\ state with $v_X = 0$ and $1$. Special efforts have been made to record ``clean'' LIF spectra, in which one progression is strongly dominating, so often several possible excitations of given ($v',J'$) from different $X$-state levels have been tried. To achieve a reasonable signal-to-noise ratio, the typical acquisition time was about 4 hours.

\begin{figure*}
  \centering
  \epsfig{file=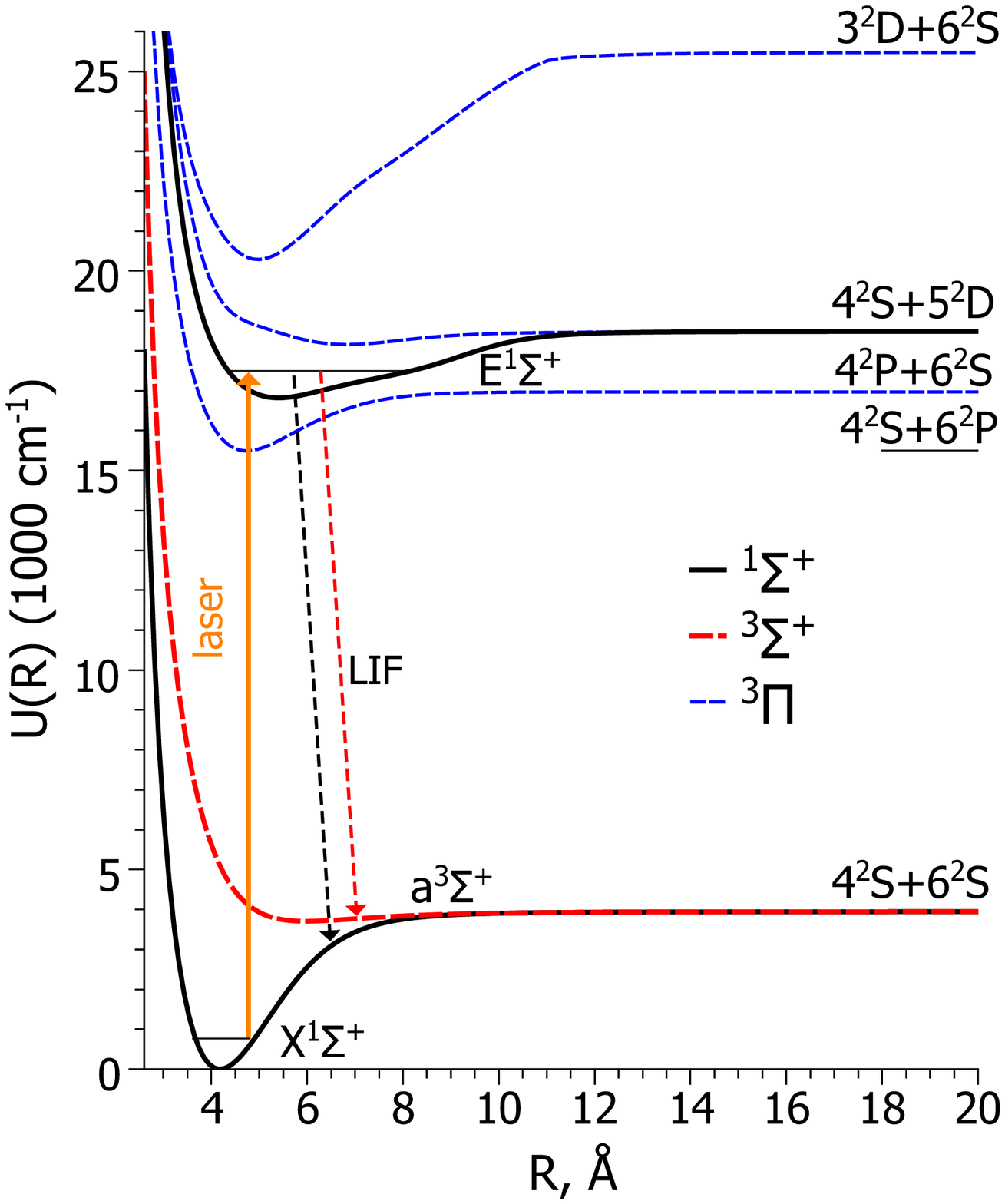,width=0.33\linewidth}(a)
  \epsfig{file=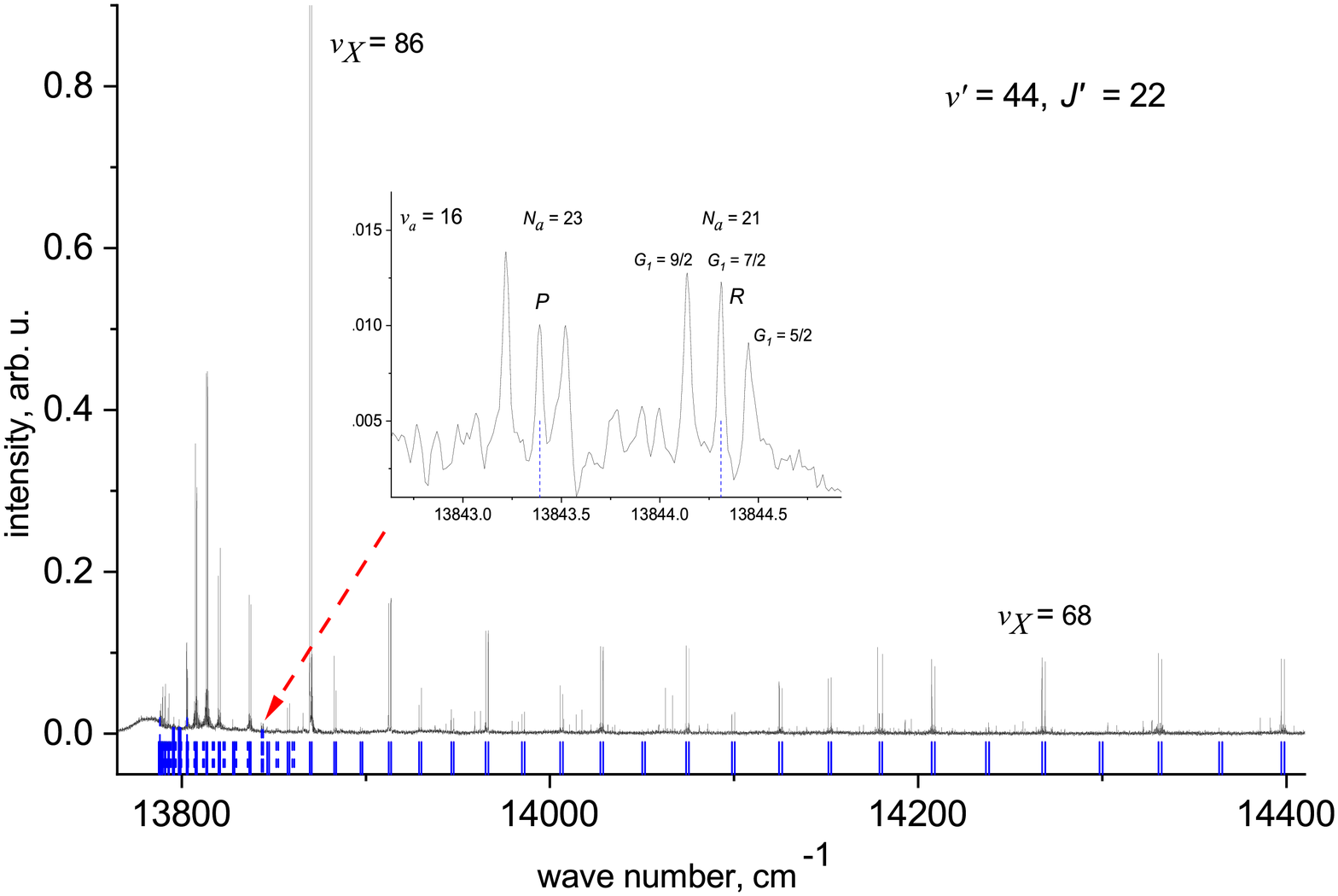,width=0.6\linewidth}(b)
 \caption{(a) Scheme of the scalar-relativistic electronic terms of KCs \cite{Korek:2000} relevant for the present study. (b) Example of LIF $E(4)^1\Sigma^+\to (a^3\Sigma^+,X^1\Sigma^+)$ spectrum containing progressions originating from $v^\prime$ = 44, $J^\prime$ = 22 level with energy 17857.423 \cmm; the spectral resolution is set 0.03 \cmm. Vertical blue solid lines below the spectrum mark the calculated positions of transitions (with the HFS deperturbation model from \cite{Krumins:2022}) to $v_X$ and dotted lines to $v_a$. Dotted red arrow marks the position of $P,R$ transitions to $v_a$ = 16, which HFS is shown in the inset. The triplet HFS components are labeled with their $G_1$ values (see Appendix~\ref{HFS_CC} for details).}
\label{Spec1}
\end{figure*}

As established in \cite{Ferber:2013}, the excitation of \Estate\ state levels with $v^\prime = 44$ and $45$ and various $J^\prime$ values within 6 -- 22 results in fluorescence progressions to the \Xstate\ ($v_X$, $J_X$) and \astate($v_a$, $J_a$) states, stretching to the highest values of $v_X$ and $v_a$ close to the dissociation limit, which are strongly coupled by HF interaction, see Fig.~\ref{Spec1}a.

Fig.~\ref{Spec1}b presents a part of $E$ ($v^\prime$ = 44, $J^\prime$ = 22) $\rightarrow$ ($a,X$) LIF spectrum, with a strong $P$, $R$ doublet progression to the $X$ state proceeding up to the dissociation limit.  As can be seen, the strongest $E - X$ line is observed for $v_X$ = 86. Since the optical filter cuts the LIF to lower $v_X$ our analysis of intensity distribution starts from $v_X$ = 68 for most of progressions. The inset zooms in a fragment of the spectrum, in which the $E - a$ transition to $v_a$ = 16 is also visible. One can recognize the characteristic three groups of lines, which appear due to the HFS splitting of the $a^3\Sigma^+_{1}$ $e$ levels by the magnetic dipole interaction \cite{Kato:1993, Busevica:2011, Ferber:2013, Krumins:2022}. These triplet lines correspond roughly to the $G_1=9/2$, $G_1=7/2$ and $G_1=5/2$ components according to the Hund case coupling ($b_{\beta S}$) of the total electron spin with $\bm{S}=1$ and the nuclear spin of Cs $\bm{I}_{\mathrm{Cs}}=7/2$ to an intermediate angular moment $\bm{G_1}$=$\bm{S}$+$\bm{I}_{\mathrm{Cs}}$ (see Appendix~\ref{LinInt} and~\ref{HFS_CC}). The additional splitting of these triplet HFS components caused by a nuclear spin of K atom was completely ignored hereafter since it could not be resolved in the Doppler-limited LIF spectra. Appearance of singlet-triplet transitions to the almost pure \astate\ lines can only be explained by the entire non-vanishing $E - a$ transition dipole moment function, because as already shown in \cite{Busevica:2011, Szczepkowski:2012} the \Estate\ state is free from local spin-orbit perturbations (see also Fig~\ref{Spec1}a and Appendix~\ref{Ea} for details).

\begin{figure}
  \centering
  \epsfig{file=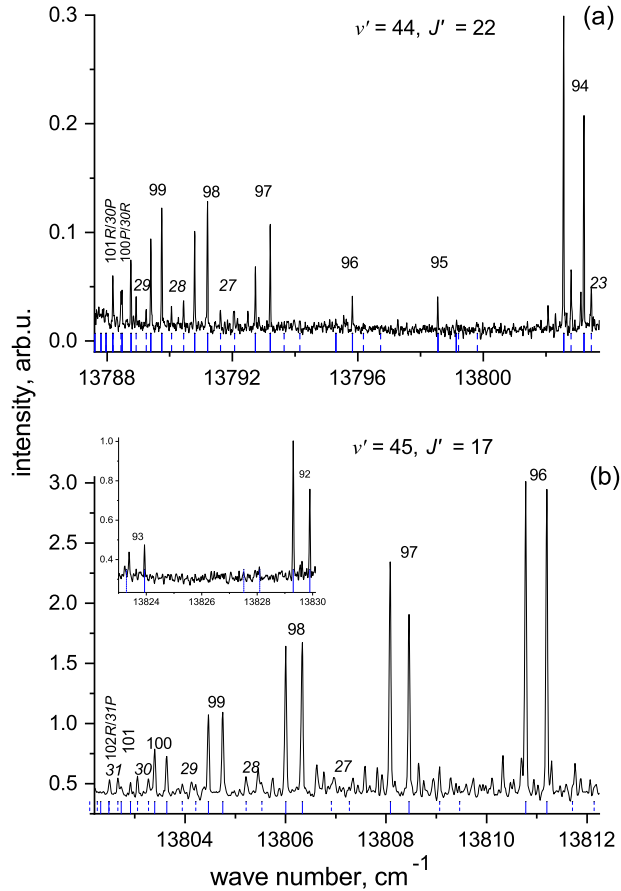,width=0.5\linewidth}
 \caption{Examples of LIF spectra. (a) The low frequency part of the progression shown in Fig.~\ref{Spec1}. Numbers denote vibrational levels ($v_a$ -numbering is in italic). The visually overlapping lines 100$P$/30$R$ can be resolved at 0.02 \cmm\ resolution. The changeover of $P$, $R$ doublet lines intensities at transitions to $v_X$ = 95 and 96 is clearly seen. (b) The very end of a progression from the excited level $v^\prime$ = 45, $J^\prime$= 17 with energy 17871.665 \cmm. The inset zooms in a changeover of $P$, $R$ doublet lines intensities at $v_X$ = 92 and 93.}
\label{Spec2}
\end{figure}

The high $v_X$-range of the same spectrum is given in Fig.~\ref{Spec2}a. Strong local $X-a$ perturbation may be recognized for $v_X$ = 94 due to appearance of extra lines. They are explained by the HF-mixing of the $v_X$ level with the triplet state level with $v_a$ = 23. Please note, that now the HFS of the \astate\ levels does not show up, but only the central $G_1=7/2$ component, which is involved in the $X-a$ interaction. Such $E-a$ $P, R$ doublets, without HF splitting, appear systematically close to the dissociation limit starting from levels above $v_a$ = 27, which have singlet admixture. Another interesting feature is the changeover in the relative intensities of $P$ and $R$ components with $v_X$ = 95 (dominating $P$-component) and 96 (dominating $R$-component). Fig.~\ref{Spec2}b presents the low frequency end of another pair of LIF progressions starting from $v^\prime = 45$, $J^\prime = 17$ level, in which one also can see the $E-a$ doublets progression above $v_a=27$. The inset illustrates the changeover between the intensities of $P$ and $R$ components of the $E-X$ progression for $v_X$ = 92 and 93.

\begin{figure}
  \centering
  \epsfig{file=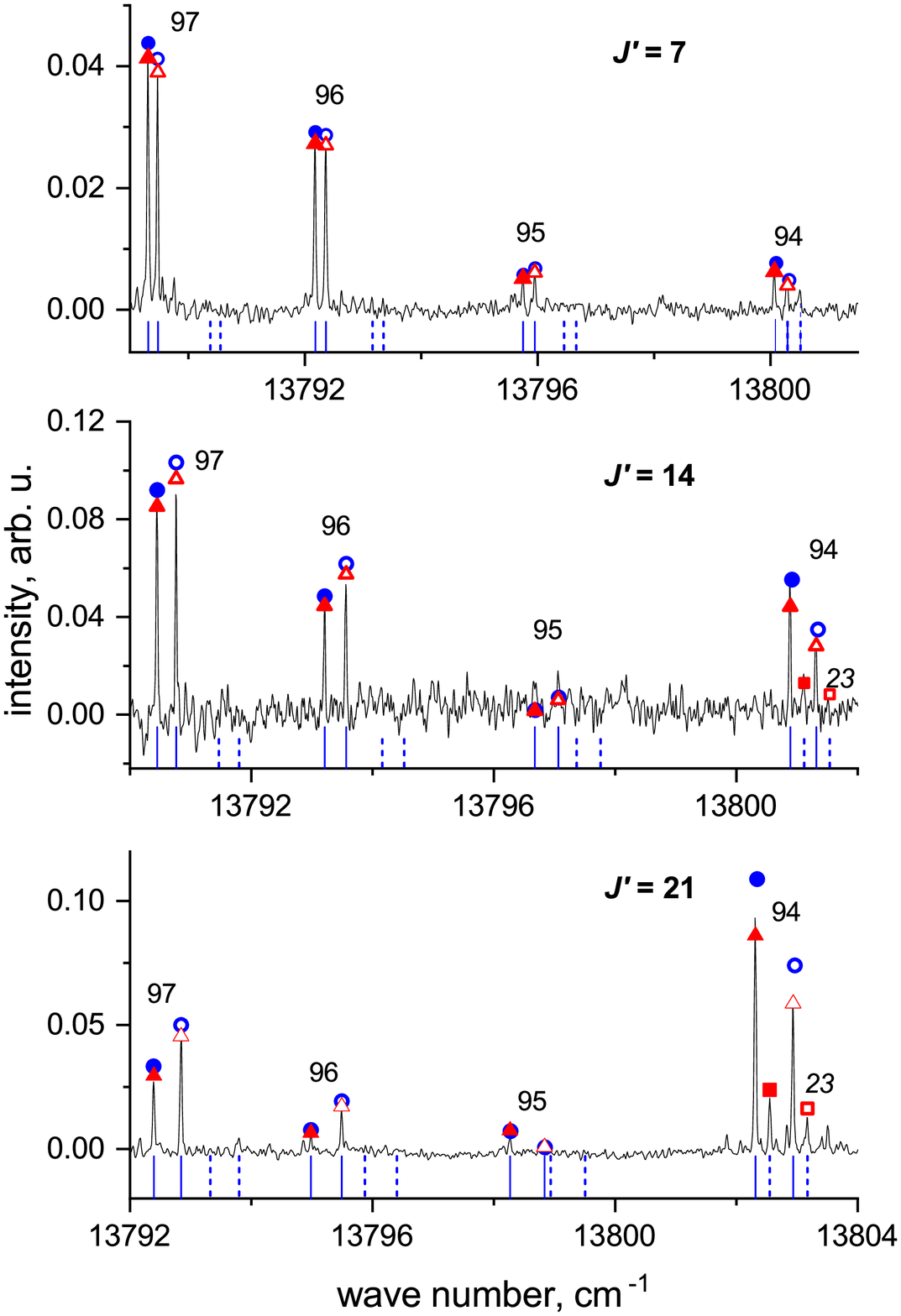,width=0.5\linewidth}
   \caption{Fragments of spectra with $E-X$ progressions starting from $v^\prime = 44$ levels with $J^\prime$ = 7, 14, and 21 for $v_X$ ranging from 94 to 97, which demonstrate different behavior of the relative line intensities in $P/R$ doublets. Symbols indicate calculated and scaled intensities with (red triangles) and without (blue circles) accounting for the HF interaction between $X$ and $a$ states. Red squares mark the $E-a$ transition to $v_a=23$, which appears due to the local HFS interaction with $v_X=94$.}
 \label{Change}
\end{figure}

Fig.~\ref{Change} presents fragments of $E-X$ LIF progressions from $v^\prime = 44$ with $J^\prime$ = 7, 14, and 21, which have been chosen to demonstrate the characteristic features related to rotational-vibrational interaction. Thus, for $v_X= 97$ the changeover between the intensities of the $P$ and $R$ components is visible when comparing LIF from $J^\prime = 7$ and $14$, while for $v_X = 95$ the change of the relative intensities takes place at somewhat higher $J^\prime$ -- compare spectra from  $J' = 14$, $J'=21$ and also $J^\prime = 22$ (Fig.~\ref{Spec2}a).

Detailed analysis of the intensity distribution of recorded LIF $E – (a,X)$ progressions from $v' = 44$ and $45$ with various $J'$ values was carried out. Line intensities were measured from the peak of the line profile. It was necessary to accurately account for the spectral background, as well as for possible contributions of overlapping satellite lines. Such lines appear due to population of neighboring $E$ state rotational levels by collisions with the atoms of the inert buffer gas. First, we determined relative intensities of satellite lines with respect to the optically excited ``main'' line and for this we used the $R$-component of the strongest transition to $v_X$ = 86 for $v' = 44$ and to $v_X = 87$ for $v'= 45$, where the satellites could be clearly resolved.  Then we simulated the relaxation line patterns through the whole spectrum, since their widths and frequencies as well as relative intensities with respect to the ``main'' line are known. Finally, the contribution of relaxation was subtracted from the signal of lines under study and their intensity was determined as difference between the peak value and the corrected background. It should be noted that the lines of $E\to a$ LIF progressions to low and intermediate $v_a$ values show triplet HFS (see the inset in Fig~\ref{Spec1}b). In these cases only the intensity of the central, $G_1=7/2$ component is used in the analysis.

Determination of LIF intensity distribution in the actual spectra involves also the issue of calibration of the overall spectral sensitivity of the detection system (optical filter, beam splitter, photodetector etc.). In the present analysis we used the spectral sensitivity curve S($\nu$) determined in \cite{Klincare:2012}. The overall correction was about 8 percent between the ends of spectral range. Each progression was recorded several times and then averaged. This allowed us to estimate the statistical uncertainty for the experimental intensities as being no more than 5 $\%$ for the lines with signal-to-noise ratio (SNR) above 10.

\section{Outline of the $E\to (a,X)$ intensity simulations}
\label{theory:sec}

In this section we provide only main details on the simplified non-adiabatic model which was used for explaining the relative intensity distribution of $E\to (a,X)$ LIF progressions. The full computational procedure implemented in a code for transition probabilities is given in Appendix~\ref{LinInt}. 

The absolute line intensity [W/m$^2$] is proportional to:

\begin{equation}\label{Int1}
   I^{Calc}_{E\to a,X}\sim \nu_{E-(a,X)}^4|M_{E-(a,X)}|^2 \mbox{ ,}
\end{equation}

\noindent where $\nu_{E-(a,X)}=E_E(v_E^{\prime},J^{\prime})-E_{(a,X)}(J''=J^{\prime}\pm 1$) is the wave number of the transition, and

\begin{equation}
M_{E-(a,X)} \approx \alpha_{\parallel}(J^{\prime})\langle v^{J^{\prime}}_E|d_{EX}|\xi^{J''}_X\rangle
+ \alpha_{\perp}(J^{\prime})\langle v^{J^{\prime}}_E|d_{Ea}|\xi^{J''}_a\rangle
\label{Int2}
\end{equation}

\noindent is the rovibronic matrix element of the corresponding transition dipole moment (see Appendix~\ref{LinInt}). Only the $e$-symmetry $F_2$ components of the \astate state (with $N_a=J^{\prime}\pm 1$) are involved due to the $e$-symmetry of the \Estate state levels (see Appendix~\ref{HFS_CC} and also Ref.~\cite{Krumins:2022}). $d_{EX}(R)$ and $d_{Ea}(R)$ are electronic transition dipole moments, which are available as functions of the internuclear distance $R$ from \emph{ab initio} electronic structure calculations~\cite{Klincare:2012}. 

The adiabatic rovibronic energies of the ``shelf-like'' \Estate state and the corresponding rovibrational wave functions $|v^{\prime}_E(J^{\prime})\rangle$ were obtained from the empirical potential $U_E(R)$~\cite{Busevica:2011, Szczepkowski:2012}. The conventional adiabatic approximation has been applied for the isolated state although this singlet state is supposed to undergo regular spin-orbit mixing with the remote $^3\Pi$ states manifold (see Fig.~\ref{Spec1}a) introducing a non-zero singlet-triplet $d_{Ea}(R)$ transition moment (see details in Appendix~\ref{Ea}).

In turn, the non-adiabatic eigenvalues $E_{(aX)}$ and $R$-dependant mixing coefficients $|\xi_{(aX)}\rangle$ of the two-component vibronic eigenfunction

\begin{equation}\label{Psi_Xa}
\Psi_{(aX)} = \Phi^{el}_X|\xi_X\rangle + \Phi^{el}_a|\xi_a\rangle
\end{equation}

\noindent corresponding to the bound levels of the mutually perturbed \Xstate\ and \astate\ states were obtained in the framework of the CC deperturbation model briefly discussed in Appendix~\ref{HFS_CC} and in details in Ref.~\cite{Krumins:2022}. The components of the vibrational wave function $|\xi_X\rangle$ and $|\xi_a\rangle$ in Eq.~(\ref{Psi_Xa}) are normalized as $P_X$+$P_a$=1, where $P_i=\langle\xi_i|\xi_i\rangle_R$ with $i\in$[\Xstate,\astate]. It should be reminded that the conventional labeling of levels by vibrational quantum numbers $v_a$ and $v_X$ is not exactly correct for the non-adiabatically coupled states since their nodal structure does not satisfy the condition of the \emph{oscillation theorem}~\cite{Landau}. Nevertheless for the sake of simplicity we use it throughout the paper to indicate the predominant character of the wave function. In fact the dominant term $P_a$ or $P_X$ for most of the observed levels was always above $0.7$ and reached about $0.5$ only for the highest vibrational levels.

The two-by-two CC system of radial equations accounting explicitly for the hyperfine magnetic dipole interaction between the \Xstate\ state and the $\Omega=1$ component of the \astate\ state was numerically solved by the Fourier-Grid-Hamiltonian method~\cite{FGH} (see Appendix~\ref{HFS_CC}). To account correctly even for the levels close to the dissociation limit, the grid was composed out of 901 points from 2.2 \AA\ to 40 \AA. We computations were performed with the same code as in Ref.~\cite{Havalyova:2021}, supplied with a new module for evaluation of rovibronic transition probabilities.

The $M$-independent direction cosine matrix elements $\alpha_{\parallel/\perp}(J^{\prime})$ involved in Eq.(\ref{Int2}) correspond to $3j$ symbols in Eq.(\ref{eq:line_strength2}), and their expressions can be found for all rotational branches in closed analytical form (see, for example, Table 6.1 of Ref.~\cite{Field} and also Appendix~\ref{LinInt}) for both $\Omega^{\prime}=\Omega^{\prime\prime}$ (the so called \emph{parallel}) and $\Omega^{\prime}=\Omega^{\prime\prime}\pm 1$ (\emph{perpendicular}) electronic transitions. The squares of the rotational matrix elements $\alpha_{\parallel/\perp}(J^{\prime})$ correspond to the well-known H\"onl-London factors~\cite{Field}, which were normalized~\cite{Warson2008} as $\sum_{J^{\prime\prime}} |\alpha_{\parallel/\perp}|^2 = (2J^{\prime}+1)(2S+1)$, where $S=0,1$ for the $E\to X$ and the $E\to a$ transitions, respectively.

It should be reminded that the $\alpha_{\parallel}$ and $\alpha_{\perp}$ values and $J^{\prime}\rightarrow J+1$ have opposite signs \cite{Kato:1993} while their absolute values rapidly approach each other for both $J^{\prime}\rightarrow J\pm 1$ branches as the rotational quantum number $J^{\prime}$ increases: $|\alpha_{\parallel}|\approx |\alpha_{\perp}|\approx J^{\prime}$. In this case, the relation (\ref{Int2}) can be reduced to the form

\begin{equation}
\label{Int_approx}
|M^{P/R}_{E-(a,X)}|\sim|M^{P/R}_{EX}\pm M^{P/R}_{Ea}| \mbox{ ,}
\end{equation}

\noindent where

\begin{equation}
\label{M_EXa}
M^{P/R}_{EX}=\langle v^{J^\prime}_E|d_{EX}|\xi^{J^\prime \pm 1}_X\rangle,\quad
M^{P/R}_{Ea}=\langle v^{J^\prime}_E|d_{Ea}|\xi^{J^\prime \pm 1}_a\rangle \mbox{ .}
\end{equation}

Using explicit relations for rotational $\alpha_{\parallel/\perp}$-factors from Ref.~\cite{Field} we can write:

\begin{eqnarray}\label{PR_ratio}
\frac{I^P}{I^R}\approx \left(\frac{\nu^P}{\nu^R}\right)^4\left[\frac{\sqrt{J^\prime}M^P_{EX}-\sqrt{J^\prime+1}M^P_{Ea}}{\sqrt{J^\prime+1}M^P_{EX}+\sqrt{J^\prime}M^R_{Ea}}\right]^2 \mbox{ .}
\end{eqnarray}

For relatively high $J^{\prime}$-values the relation (\ref{PR_ratio}) can be further reduced to:

\begin{eqnarray}\label{PR_ratio1}
\frac{I^P}{I^R}\approx \left|\frac{\langle v^{J^\prime}_E|\xi^{J^\prime+1}_X\rangle}{\langle v^{J^\prime}_E|\xi^{J^\prime-1}_X\rangle}\right|^2
\left[\frac{1-K^P}{1+K^R}\right]^2
\end{eqnarray}

\noindent under well-established $R_c$-centroid approximation~\cite{Fraser}. It is clearly seen from Eq.(\ref{PR_ratio1}) that the sign and the absolute value of the ``sensitivity'' coefficients

\begin{eqnarray}\label{K_PR}
K^{P/R} = \frac{d_{Ea}(Rc_a)}{d_{EX}(Rc_X)}\frac{\langle v^{J^\prime}_E|\xi^{J^\prime\pm 1}_a\rangle}{\langle v^{J^\prime}_E|\xi^{J^\prime\pm 1}_X\rangle};\qquad
Rc_{(a,X)} = \frac{\langle v^{J^\prime}_E|R|\xi^{J^\prime\pm 1}_{(a,X)}\rangle}{\langle v^{J^\prime}_E|\xi^{J^\prime\pm 1}_{(a,X)}\rangle}
\end{eqnarray}

\noindent should be responsible for the most part of peculiarities observed in the $P/R$ intensity ratios as a function of vibrational and rotational quantum numbers of upper and lower electronic states. The most pronounced variation of the $I^P/I^R$-values are expected for rovibronic transitions when $|K^{P/R}|\approx 1$ and the overlap integrals $\langle v^{J^\prime}_E|\xi^{J^\prime\pm 1}_X\rangle$  are small enough~\cite{FCF_J}.

The correct comparison of the calculated $E\to (a,X)$ line intensities $I^{\mathrm{Calc}}$ with their measured counterparts $I^{\mathrm{Expt}}$ as registered by the detector depends on a series of additional factors (population of the initial state, spectral sensitivity of the detector and the transmission of the optics, geometry of observation, etc.), which are not always easy to account for. Therefore it is much easier to compare relative intensities of spectral lines (corrected for the spectral response of the detecting system) which share common upper $E$-state level since most of these factors are canceled. In the present study we normalized all theoretical and experimental intensities of a given $E \to (a,X)$  progression to the intensity of the strongest among the observed lines. As already mentioned, the $R$ component of the transition to $v_X=86$ (for $v'=44$) or to $v_X=87$ (for $v'=45$) were chosen, because they are well isolated from the rotational relaxation satellites.

\section{Results}
\label{res:sec}

In this Section we present typical examples, where the experimental SNR allows to recognize most of the theoretically predicted lines. We selected a high $J'$ and a low $J'$ progression from $v'=44$ and one progression from $v'=45$. It this way we can cover virtually all interesting phenomena and also demonstrate that the applied theoretical CC approach is able to account for most of them.

\begin{figure}
  \centering
  \epsfig{file=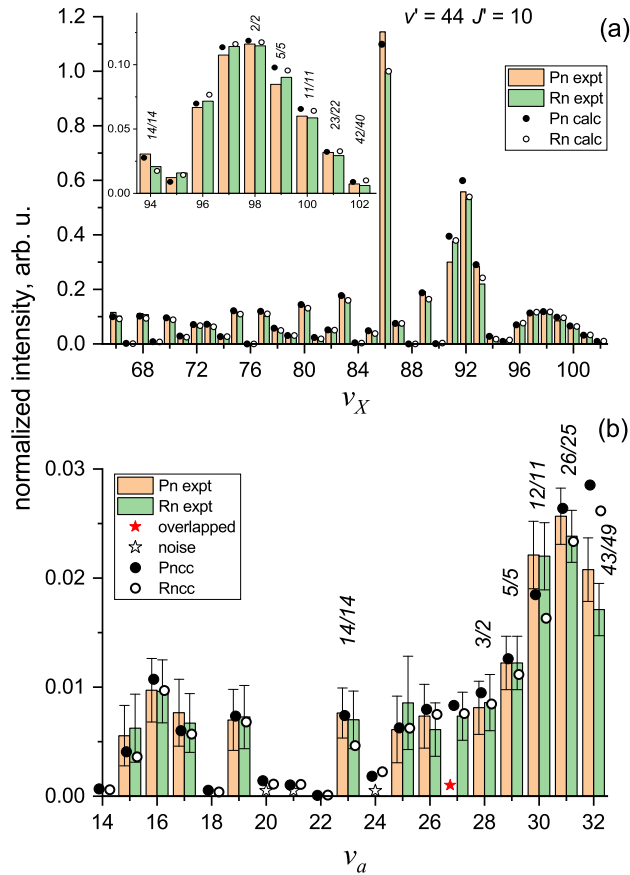,width=0.5\linewidth}
 \caption{Experimental and calculated relative intensity distributions in $E-(a,X)$ LIF progressions starting from $v^\prime$ = 44, $J^\prime$ = 10 level. Vertical bars are the experiment, circles – CC calculations. All values are normalized to the $E-X$ $R$ transition with $v_X$ = 86. Notations $Pn$ and $Rn$ relate to the normalized values. The red asterisk marks the overlapping lines. Open asterisks stay for lines, which can not be distinguished from noise level. Error bars are the estimated experimental uncertainties. Above the bars the mixing coefficient (in \%) is given for selected $v_X$ and $v_a$ $P/R$ components, see text for details.}
 \label{4410}
\end{figure}

In Figure~\ref{4410} we compare the experimentally determined intensities with the results from the CC code. The upper level is \Estate\ ($v'=44, J'=10$). Relative intensities are shown, scaled with the intensity of $E(v'=44,J'=10)\rightarrow X(v_X=86, J_X=9)$ transition. The normalized partial components $P_X$ and $P_a$ (see Section~\ref{theory:sec}) of the respective vibrational wave functions (in \%) are presented for selected transitions above the bars. The numbers separated with forward slash are designating the $P$ and $R$ branches. For the $E-X$ transitions the contribution due to the $a$ state is given ($P_a$), while for the $E-a$ transitions -- due to the $X$ state contribution $P_X$. For $E-X$ lines (Fig.~\ref{4410}a) the calculation reproduces correctly not only the nodal structure of the spectrum, but also quite reasonably the overall intensity distribution, including the ratio between $P$ and $R$ components (e.g. $v_X=83,86,92,93$, and $96$) and also the changeover which occurs for this $J'$ between $v_X=94$ and $95$ (see the inset in Fig.~\ref{4410}a). For $v_X=91$, however, the model predicts $P/R$ ratio, which disagrees with the experiment. It is not excluded that such sharp deviation, which exceed the typical experimental uncertainty of about 5 \%\ is caused by accidental absorption of the LIF line at $13818.612$ \cmm\ in KCs \Xstate-c$^3\Sigma^+$ \cite{Szczepkowski:2018,Kruzins:2021} or K$_2$ X$^1\Sigma^+_{\mathrm{g}}-$A$^1\Sigma^+_{\mathrm{u}}$ bands \cite{Manaa:2002}.

The overall agreement for the $E\rightarrow a$ part of the fluorescence is also very convincing (see Fig.~\ref{4410}b) taken into account the much weaker experimental intensities, which should be compared to the noise level of about 0.002 relative units for this spectrum. Therefore, for example, the transition to $v_a=24$ is missing, because its intensity is comparable to the noise. Some transitions are blended by other lines (e.g. the $P$ component to $v_a=27$), so their absence from the Figure does not demonstrate any disagreement with the experiment. The transitions up to $v_a=20$ levels are due to the nonzero transition dipole moment $d_{Ea}$. These levels are of predominantly triplet character, isolated from the $X$ state and therefore intensity borrowing due to mixing with the \Xstate\ state is impossible. For higher $v_a$ the appearance of the $E-a$ lines gradually changes from lines with HFS for ``pure'' \astate\ levels (like in the inset of Fig.~\ref{Spec1}b) and to a single line pattern after about $v_a=20$, where the mixing with the $X$ state increases (see e.g. Fig~\ref{Spec2}). Apparently the direct $E-a$ transitions ($M_{Ea}$ in Eq.~\ref{M_EXa}) vanish close to the dissociation limit and transitions to the \astate\ are mainly due to the $M_{EX}$ part of eq.~\ref{M_EXa}. The lines appear as singlets because only the $G_1=7/2$ component of the HF triplet mixes with the $X$ state. The most critical point is the intensities of lines to levels $v_X=94,95$ and $v_a=23,24$, where \Xstate\ and \astate\ states come to the closest separation and locally perturb each other. The changeover of the $P/R$ ratio between $v_X=94$ and $95$ in Fig.~\ref{4410}a is also nicely reproduced. So the CC model \cite{Krumins:2022} (based solely on experimental frequencies) not only correctly calculates the level energies, but also finds correctly the two component wave functions. The agreement between experimental and calculated weak $E-a$ intensities is even more remarkable if one takes into account that all $E-a$ and $E-X$ transitions are normalized to the same singlet transition. A correct ratio between singlet and triplet parts of the spectra means that  correct are also the relative values of the $d_{EX}(R)$ and $d_{Ea}(R)$ dipole moments.

\begin{figure}
  \centering
  \epsfig{file=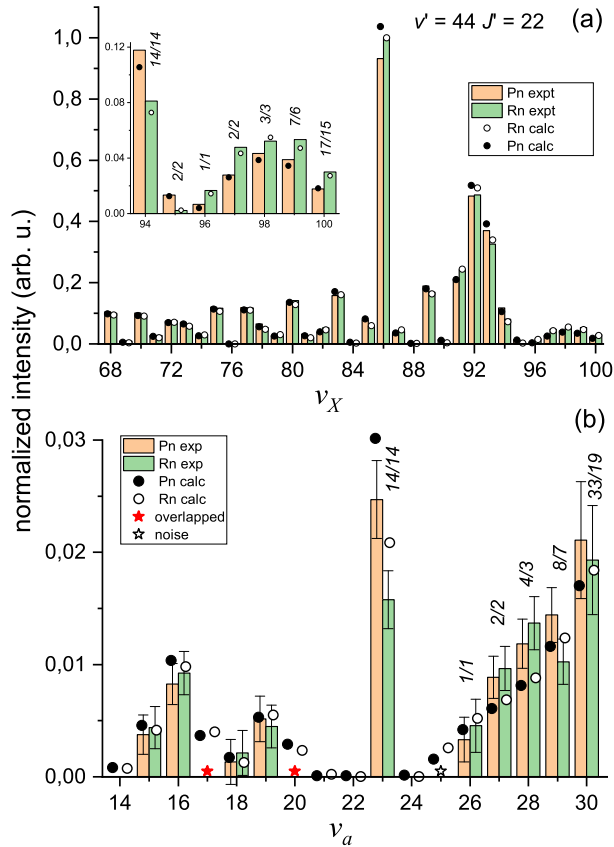,width=0.5\linewidth}
   \caption{Experimental and calculated relative intensity distributions in $E-(a,X)$ LIF progressions starting from $v^\prime$ = 44, $J^\prime$ = 22 level. Notations are the same as in Fig.~\ref{4410}.}
   \label{4422}
\end{figure}

Another example of intensity distribution is presented in Fig.~\ref{4422}, in which the upper level is  \Estate\ ($v'=44, J'=22$).  Similarly to the case with $J’ = 10$ (Fig.~\ref{4410}), both measured and calculated intensities are normalized to the $E(v'=44,J'=22)\rightarrow X(v''=86, J''=21)$ ``clean'' $R$ line intensity. Fig.~\ref{4422}a demonstrates excellent agreement between theoretical and measured values, except for $v_X = 86$ where the experimental value of $P$ component, contrary to calculations, is by about 10\%\ smaller than that of $R$ component. In the inset of the figure one can see that the intensity changeover between $P$ and $R$ lines, which in this case appears for $v_X=95$ and $v_X=96$ is nicely described by  theory. Similar changeover for $v_X=95$ and $v_X=96$ was observed in the case of progression with $J’=21$, see Fig.~\ref{Change}. Note that the changeover for $J’= 22$ is  more pronounced than for $J’= 10$, see Fig.~\ref{4410}a. Presented in Fig.~\ref{4422}b intensity distribution to $a$ state also shows good agreement between experiment and theory.

\begin{figure}
  \centering
  \epsfig{file=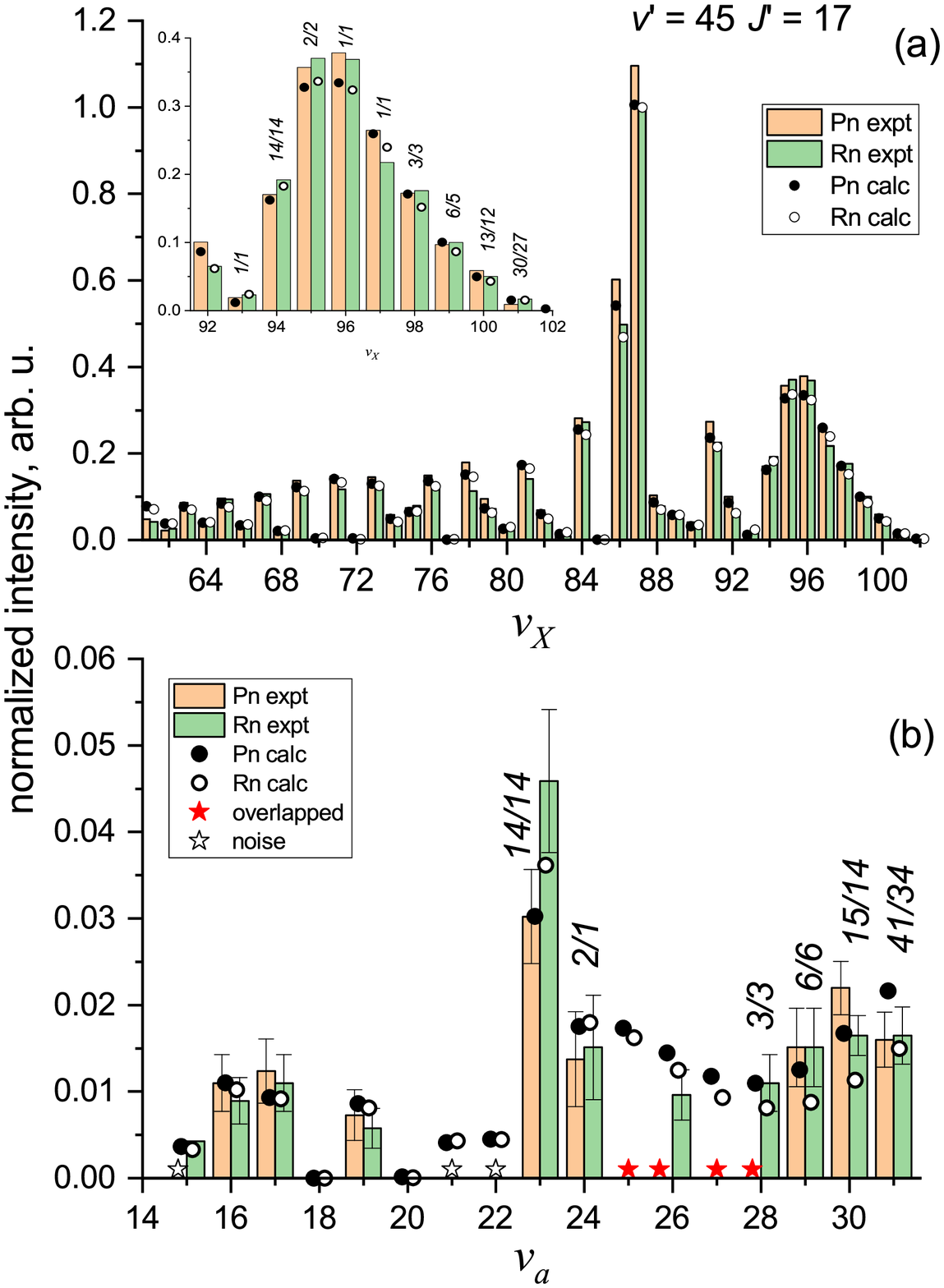,width=0.5\linewidth}
   \caption{Experimental and calculated relative intensity distributions in $E-(a,X)$ LIF progressions starting from $v^\prime$ = 45, $J^\prime$ = 17 level. All values are normalized to the $E-X$ transition to $v_X$ = 87 ($R$ component). The notations are the same as in Fig.~\ref{4410}.}
 \label{4517}
\end{figure}

In Fig.~\ref{4517}a we present the intensity distribution for $E-X$ progression from a higher $E$ state level ($v'=45, J'=17$). Both measured and calculated intensities are normalized to the strongest  $E(v'=45,J'=17)\rightarrow X(v_X=87, J_X=16)$ line intensity. In a broad scope this picture confirms the quality of the CC model. Note that a change of upper state vibrational level causes a shift of the changeover range to $v_X=92$ and $v_X = 93$, which is again correctly reproduced by the model. As to the $E-a$ transitions, see Fig.~\ref{4517}b, we should draw reader’s attention to a different intensity distribution at $v_a > 23$ than for case of $v’= 44$. Despite of sparse experimental data, these observations in general prove the quality of the calculated dependence.

\begin{figure}
  \centering
  \epsfig{file=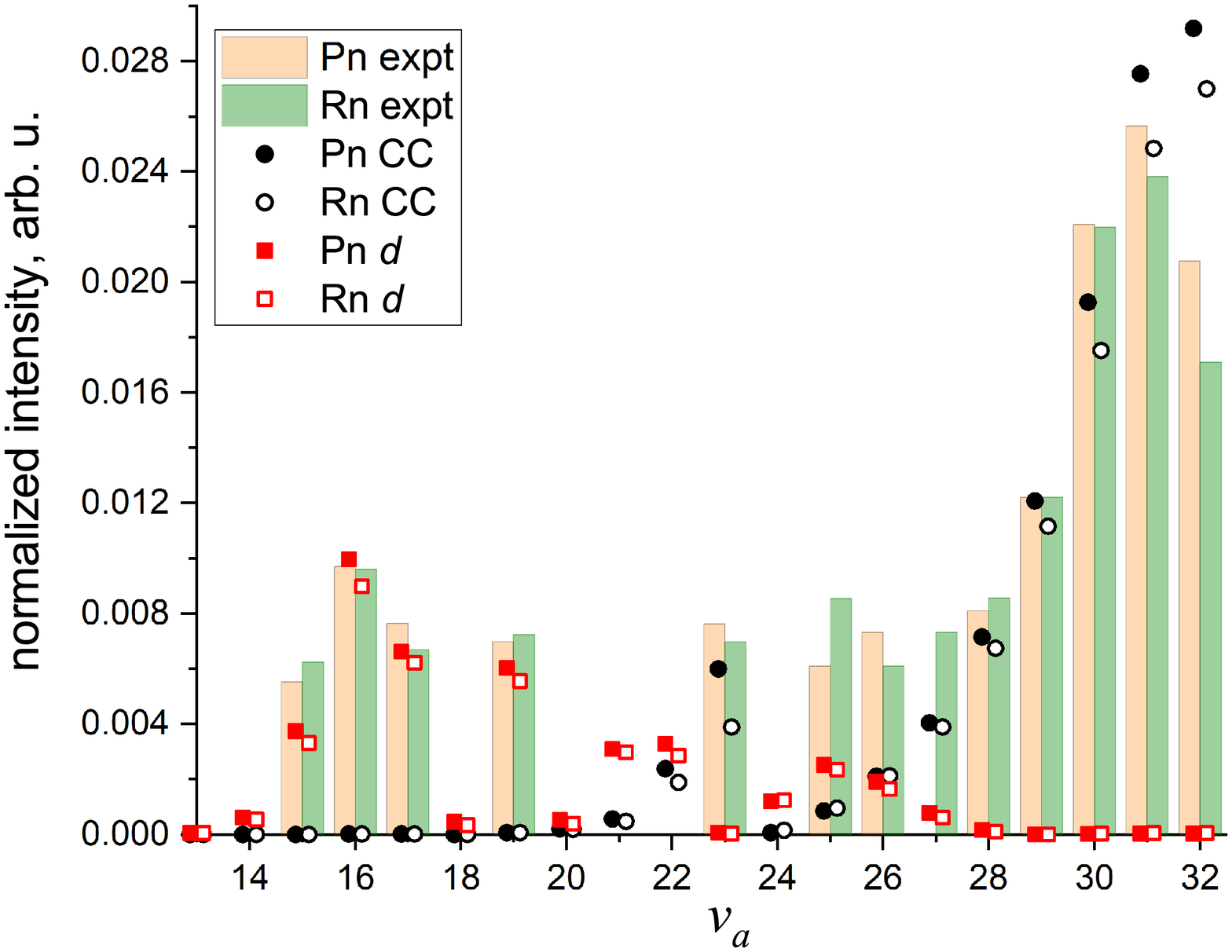,width=0.5\linewidth}
 \caption{Line intensities from the $E$ state $(44,10)$ level with CC calculation but $d_{Ea}=0$ (black dots) and  without CC calculation but nonzero $d_{Ea}$ (red squares). Both calculations can not reproduce the experimental data (bars) for all $v_a$.}
 \label{4410m}
\end{figure}

To demonstrate separately the contributions of line intensity borrowing by the HF mixing and the $d_{Ea}$ dipole moment in the $E-a$ transitions in Figure~\ref{4410m} we present a calculation of the line intensities from the $E$ state $(44,10)$ level with CC calculation when $d_{Ea}$ is set to zero ($I_{E-a}\sim M_{EX}^2$, black dots) and  without CC calculation and nonzero $d_{Ea}$ ($I_{E-a}\sim M_{Ea}^2$, red squares). The experimental intensities are drawn with bars. One should compare these results with the calculations in Fig.~\ref{4410} where both HF coupling and $d_{Ea}(R)\neq 0$ were taken into account. One may see that, as expected, below $v_a=20$ the role of $X-a$ mixing is negligible, whereas the presence of $d_{Ea}$ alone cannot explain the intensity distribution above $v_a=20$. In the intermediate region between $v_a=20$ and $v_a=26$ both contributions are important. It is curious to compare the intensities for $v_a=22$ and $v_a=26$. For these levels both $M_{EX}$ and $M_{Ea}$ values predict comparable intensities when taken separately. However their joint effect results in a nearly zero intensity for $v_a=22$ and a significant one for $v_a=26$ (see Fig.~\ref{4410}). The reason is that for $v_a=22$ $M_{EX}$ and $M_{Ea}$ have different signs, while for $v_a=26$ the signs are the same.

We would like to mention one more detail on the changeover of the $P/R$ ratio in the vicinity of \Xstate\ --     \astate\ perturbation (see Fig.~\ref{Change}). Initially our guess was that here we observe interference effects (see chapter 6.3 of Ref.~\cite{Field}) because all necessary conditions were fulfilled: parallel and perpendicular transitions contribute to the line intensity and therefore the relative signs for $M_{EX}$ and $M_{Ea}$ change for $P$ and $R$ transitions. However the calculated intensities with and without HF coupling  (blue and red open circles in Fig.~\ref{Change}) show that in fact the changeover in this case is a pure adiabatic rotational (centrifugal distortion) effect while the non-adiabatic HF coupling influences only the overall intensity and leaves the $P/R$ ratio almost unchanged. Apparently, the mixing between interacting states (see the values of $P_a$ and $P_X$ in Figs.~\ref{4410} and \ref{4517}) is not sufficient in order to make the values of $M_{EX}$ and $M_{Ea}$ comparable for the $E-X$ transitions with predominantly singlet character. Interference effects might be observed in the $E-a$ transitions, but unfortunately these lines are weak and in present study the possible effects are most likely hidden within experimental uncertainty.

\section{Concluding remarks}

In the present study it was shown that by constructing a realistic coupled-channels model and taking into account all important interactions it is possible to reproduce the experimental line intensities even in such a non-trivial case like the \Estate$\to$(\astate,\Xstate) band system. The model functions were either empirical PECs (\cite{Krumins:2022,Busevica:2011,Szczepkowski:2012}) or {\it{ab initio}}	 coupling functions (the HF $A(R)$ function, Appendix~\ref{HFS_CC}) and transition dipole moments $d_{EX}$ and $d_{Ea}$~\cite{Klincare:2012}. It is difficult to estimate which of these functions has more impact on the accuracy of the model. In our case, lower states PECs and $A(R)$ functions were already tested in Ref.~\cite{Krumins:2022} and they reproduce accurately a huge set of experimental energies of \astate\ and \Xstate\ states, including the near asymptotic ones. Therefore it should be assumed that the wave functions of the mixed $(a,X)$ complex are accurately known. For the transition dipole moments, however, there is no additional independent experimental test to assure their quality, so here we relied entirely on the results of theoretical calculations and the results proved to be very satisfactory. In this connection, we should stress on the increasing importance of accurate theoretical calculations. So far, transition dipole moments along with radial functions for various coupling operators can not be determined unambiguously from experimental observations, unlike the PECs. In principle, of course, the $d_{EX}(R)$ and $d_{Ea}(R)$ functions could have been fitted if systematic deviations between measured and calculated relative intensities were observed, but in our case it was not necessary. The question of using the measured line intensities together with the line frequencies as experimental data is still open, and studies like this may be used as test grounds to develop stable fitting routines.

One of the keys for the success of the present modeling is the thoroughly processing of experimental spectra in order to extract the correct line intensities. As already explained, part of the lines were overlapped with rotational relaxation satellites and accounting for them was a tedious procedure, which required {\it{a priori}} knowledge of their line frequencies.

We believe that the simple CC model developed in Ref.~\cite{Krumins:2022} for KCs \astate\ and \Xstate\ states can be applied for other alkali diatomics and similar analysis of line intensities can be carried out for optimization of photoassociation or/and STIRAP processes. Of course, in some cases the upper state in cold-collisions studies is not as simple as here (a single $^1\Sigma^+$ state), but may be a mixture of singlet and triplet states (e.g. ($^1\Sigma^+,^3\Pi$) or even ($^1\Pi,^3\Sigma^+,^3\Pi$)). This however leads only to a more complex model, but a multi-component wave function for the coupled excited states can be calculated in a similar manner.

\section{Acknowledgments}

Riga team acknowledges the support from the Latvian Council of Science, project No. lzp-2018/1-0020: ``Determination of structural and dynamic properties of alkali diatomic molecules for quantum technology applications'' and from the University of Latvia Base Funding No Y5-AZ27. IH and AP acknowledge partial support from the Sofia University grant 80-10-76/15.04.2020. IH acknowledges the support from Bulgarian national program ``Young Scientists and Postdoctoral Research Fellows 2022'' of the Ministry of Education and Science. AP acknowledges partial support from BG05M2OP001-1.002-0019:``Clean technologies for sustainable environment -- waters, waste, energy for circular economy'', financed by the Operational programme ``Science and Education for Smart Growth 2014-2020'', co-financed by the European union through the European structural and investment funds. The Moscow team is grateful for the support by the Russian Government Budget (section 0110; Project Nos. 121031300173-2 and 121031300176-3).

\appendix

\section{Computational procedure for calculation of line intensities}
\label{LinInt}

In order to determine the relative line intensities for the transitions from the \Estate\ to the coupled  (\astate,\Xstate) system first we will compute the Einstein coefficients through the formula:

\begin{equation}
    A_{if} = \frac{8 \pi^{2} \nu^{3}_{if} S_{if}}{3\varepsilon_{0} \hbar c^{3} g_{i}} = 3.1361891 \times 10^{-7} \times \nu^{3}_{if} \times \frac{S_{if}}{g_{i}} \mbox{ ,}
	\label{Ein}
\end{equation}

\noindent where $i$ and $f$ are labels for the initial and the final state. In the second part of (\ref{Ein}) the numerical factor is calculated by assuming that $\nu$ is the transition wave number measured in cm$^{-1}$. $S_{if}$ is the line strength, which represents the square of the transition dipole matrix element summed over the degenerate initial and final magnetic sub-levels $M$. $S_{if}$ is expressed in Debye square units, assuming that the involved transition dipole moment is in Debye units. $g_{i} = 2J + 1$ is the initial state degeneracy factor. As a result, the Einstein coefficients will be in $s^{-1}$ units. The evaluated Einstein coefficients of spontaneous emission is related to the experimental LIF intensities as

\begin{equation}
    I^{exp}_{if} \sim \nu_{if} A_{if}
	\label{IEin}
\end{equation}

For a transition between an initial state $\vert i \rangle$ and a final $\vert f \rangle$ state, the line strength (originally defined in Ref. \cite{Whiting_1974}) is expanded as a sum over the associated $M^{\prime}$ and $M$ quantum numbers:

\begin{equation}
    \label{eq:line_strength1}
    S_{if} = \sum_{M^{\prime} M} \sum_{p} \left| \langle f \vert T^{k}_{p}(\boldsymbol{\mu}) \vert i \rangle \right| ^{2}
\end{equation}

\noindent where $T^{k}_{p}(\boldsymbol{\mu})$ are the spherical tensor components of the electric-dipole operator expressed in a space-fixed coordinate system; $p$ is the index specifying the projection onto this system and $k$ is the rank of the tensor equal to $1$ for electric-dipole single-photon transitions as in our case. Since $\boldsymbol{\mu}$ is an internal characteristic of the molecule, it is more natural to be defined in the molecule-fixed system. We transform it using the Wigner $\mathcal{D}$ rotation matrix \cite{Brown_2003}:

\begin{equation}
    T^{k}_{p}(\boldsymbol{\mu}) = \sum_{q} {\mathcal{D}_{pq}^{k}(\omega)}^{*} T^{k}_{q}(\boldsymbol{\mu}) = \sum_{q} (-1)^{p-q} \mathcal{D}_{-p-q}^{k} (\omega) T^{k}_{q}(\boldsymbol{\mu})
\end{equation}

\noindent where $q = 0, \pm 1$ are the spherical components in molecule-fixed system and $\omega$ is a short notation for the three Euler angles. To proceed, we must explicitly specify the initial and the final statevectors in \cref{eq:line_strength1}. In Hund's case coupling (\textbf{a}) they are represented by the following set of quantum numbers: $\vert i \rangle \equiv \vert \eta \Lambda S \Sigma  \rangle \vert J \Omega M \rangle $ and $\vert f \rangle \equiv \vert \eta^{\prime} \Lambda^{\prime} S^{\prime} \Sigma^{\prime}\rangle \vert J^{\prime} \Omega^{\prime} M^{\prime} \rangle $ where $\eta$ and $\eta^{\prime}$ denote additional quantum numbers and labels required to compute the matrix element but not explicitly specified among the others. Taking into account that the Wigner matrix only acts on the rotational part of the wave function and utilizing some of its properties, we arrive at the final expression for the matrix element of the electric dipole moment:

\begin{align}
    \label{eq:int2}
    \langle \eta^{\prime} \Lambda^{\prime} S^{\prime} \Sigma^{\prime} J^{\prime} \Omega^{\prime} M^{\prime} \vert T^{1}_{p} (\boldsymbol{\mu}) \vert \eta \Lambda S \Sigma J \Omega M \rangle
    &= \sum_{q=0, \pm 1} (-1)^{p-q} \langle \eta^{\prime} \Lambda^{\prime} S^{\prime}  \Sigma^{\prime} \vert T^{1}_{q}(\boldsymbol{\mu}) \vert \eta \Lambda S \Sigma \rangle \\
    &\times
    (-1)^{M-\Omega} [(2J^{\prime}+1)(2J + 1)]^{1/2} \tj{J^{\prime}}{1}{J}{M^{\prime}}{-p}{-M} \tj{J^{\prime}}{1}{J}{\Omega^{\prime}}{-q}{-\Omega}
\end{align}

\noindent with rotational conventions adopted from \cite{Kato:1993}. The last two factors represent the Wigner $3j$-symbols.

When considering a non-adiabatic coupled-channels problem, the convenient approximate separation of this matrix element into a product of electronic, vibrational, rotational and spin factors is inappropriate because the involved basis functions are no longer eigenfunctions of the total Hamiltonian. In the Hund's case coupling $b_{\beta S}$, suitable for the set of coupled \astate\ and \Xstate\ states, the total non-adiabatic wave function can be represented as a linear superposition of nuclear-spin-electronic-spin-rotational basis functions  $\Phi_{n J s} \equiv \vert \eta \Lambda S \Sigma (\Omega=\Lambda+\Sigma) s \rangle 
\vert S I_{Cs} (G_1)\rangle \vert J \Omega M \rangle$ as:

\begin{equation}
    \Psi^{m}_{J s} = \sum_{n} \xi_{n J s}^{m} (R) \; \Phi_{n J s}
\end{equation}

\noindent where $\xi^{m}_{n J s} (R)$ are the components of the $R$-dependent (rovibrational) eigenfunction corresponding to the eigenvalue with index $m$ obtained by solving the coupled-channels system i.e. by diagonalizing the Hamiltonian matrix for a specified rotational quantum number $J$ and symmetry label $s$ for $e$- or $f$-levels. Here $n$ stands for the group of quantum numbers $\eta, \Lambda, S, \Sigma, \Omega, I_{Cs}, G_1$, designating each of the basis functions. With these wave functions the line strength for a transition between an initial state $\Psi_{i}$ ($i$ stands for $m,J,s,G_1$) and a final state $\Psi_{f}$ ($f$ stands for $m',J',s',G_1$) becomes:

\begin{gather}
    \label{eq:line_strength2}
    S_{fi} = \sum_{q} \left| \sum_{n, n^{\prime}} \big\langle \xi^{m^{\prime}}_{n^{\prime} J^{\prime} s^{\prime}} (R) \left| \boldsymbol{\mu}_{q}^{n^{\prime} n}(R) \right| \xi^{m}_{n J s} (R) \big\rangle \left[(2J^{\prime}+1)(2J + 1)\right]^{1/2} \tj{J^{\prime}}{1}{J}{\Omega^{\prime}}{-q}{-\Omega} \right|^{2}
\end{gather}

\noindent where the summation over $M$ and $M^{\prime}$ is performed taking into account the properties of the $3j$-symbols and assuming unpolarized radiation and isotropic environment. In the last equation $\boldsymbol{\mu}_{q}^{n^{\prime} n}(R)$ is the transition dipole moment function i.e. the electric dipole moment operator averaged over the electronic wave functions.

The singlet \Estate and \Xstate states formally belong to a pure Hund's case coupling (\textbf{a}) whereas the triplet \astate state should correspond to a pure Hund's case (\textbf{b}). However, we deal here only with the $a^3\Sigma^+_{1e}$ component of the triplet state, the \Xstate\ and \Estate\ states are $\Omega=0^+$ and the $E$ state is the regularly perturbed by remote $^3\Pi_{0^+}$ state (see Appendix~\ref{Ea}). Therefore, all states under consideration could be reassigned to Hund's case (\textbf{c}) states with $\Omega=0^+$ for the \Xstate and \Estate states, and $\Omega=1$ for the \astate state. Using the notation we can redefine matrix elements $\boldsymbol{\mu}_{q}^{n^{\prime} n}$ in Eq.(\ref{eq:line_strength2}) as $d_{EX}(R)$ (when $q = 0$) and $d_{Ea}(R)$ (when $q = \pm 1$), respectively. Finally a comparison with Eq.~(\ref{Int2}) shows how the $\alpha$ factors are calculated.

\section{The origin of the spin-forbidden \Estate--\astate\ transition moment}
\label{Ea}

In this Section we demonstrate that effectively the matrix elements of the \Estate\ -- \astate\ transitions can be reduced to the form $\langle v_E|d_{Ea}(R)|\xi_a\rangle$ without the need to consider  explicitly the non-adiabatic interaction of the \Estate\ state with the manifold of distant triplet states. Starting from here the rotational quantum number $J$ as index is omitted where possible for sake of clarity.

The appearance of spin-forbidden transitions can be natively understood if at least one of the considered upper or lower states is a superposition of triplet or singlet components. Let us assume that the energy isolated \Estate\ state~\cite{Busevica:2011, Szczepkowski:2012} is regularly perturbed by remote triplet $^3\Pi$ states due to spin-orbit (SO) interactions (see Fig.~\ref{Spec1}). Then, using the second-order non-degenerated perturbation theory one can write for a triplet admixture of the $E$ state:

\begin{eqnarray}\label{gammaPi1}
 \langle\Psi^p_E| =\sum_{t}\Phi^{el}_t\sum_{v_t}\frac{\langle v_E|V^{so}_{Et}|v_t\rangle_R\langle v_t|}{E_E-E_t} \mbox{ ,}
\end{eqnarray}

\noindent where the index $t$ runs over all distant $^3\Pi_{0^+}$ sub-states and $V^{so}_{Et}(R)$ are electronic matrix elements of the SO interaction. Due to the approximate vibrational sum rule \cite{Stolyarov:1994} the vibronic matrix element of the transition dipole moment from the \Estate\ state to the lower \astate\ state can be represented as

\begin{eqnarray}\label{Lambda-p-delta1}
\langle \Psi^p_E|\mathbf{\hat{d}}|\Psi_a\rangle = \sum_{t}\sum_{v_t}\frac{\langle v_E|V^{so}_{Et}|v_t\rangle_R\langle v_t|d_{ta}|\xi_a\rangle_R}{E_E-E_t}\approx \langle v_E|d_{Ea}|\xi_a\rangle_R
\end{eqnarray}

\noindent where $d_{ta}(R)$ is spin-allowed triplet-triplet electronic transition dipole moment and

\begin{eqnarray}\label{Lambda-p-sigma2}
d_{Ea}=\sum_{t}\frac{V^{so}_{Et}(R)d_{ta}(R)}{U_E(R)-U_t(R)}
\end{eqnarray}

\noindent is the effective ``spin-forbidden'' transition dipole moment from the $E$ to the $a$ state.

In this way the direct transition the singlet \Estate\ and the triplet \astate\ state may be explained as a second order effect due to mixing of the $E$ state with distant triplet states (compare with Ref.~\cite{Field}, chapter 6.4). Similar spin-forbidden transitions were reported within the \Xstate\ -- c$^3\Sigma^+$ band of the same molecule observed in absorption \cite{Szczepkowski:2018, Kruzins:2021}. The lower part of the c$^3\Sigma^+$ state is free from local singlet-triplet perturbations and this transitions in a similar manner (compare with (\ref{Lambda-p-sigma2})) could be modeled by estimating the effective, second order $d_{cX}(R)$ dipole moment by summing over the interactions between the c$^3\Sigma^+$ state and the distant $^1\Pi$ states.

According to Eq.(\ref{Lambda-p-sigma2}) the sum-over-states transition dipole moment $d_{Ea}(R)$ as a function of $R$ was calculated in Ref.~\cite{Klincare:2012}. The required adiabatic potentials, singlet-triplet SO coupling matrix elements and triplet-triplet transition dipole moments were obtained in the framework of scalar-relativistic electronic structure calculations. A very similar $d_{EX}(R)$ and $d_{Ea}(R)$ functions have been evaluated directly using a fully relativistic \emph{ab initio} approach (e.g. \cite{Kruzins:2021, Oleynichenko:2021}). At the present work the required spin-allowed $E - X$ and spin-forbidden $E - a$ electronic transition dipole moment functions were borrowed from Ref.~\cite{Klincare:2012}.

\section{The \Xstate\ -- \astate\ hyperfine interaction model}
\label{HFS_CC}

In this section we remind details on the simplified coupled-channel (CC) HFS deperturbation model used~\cite{Krumins:2022} to calculate the energy levels and the wave functions of the mutually perturbed  \Xstate\ and \astate\ states.

It is assumed that the HFS of the \astate\ state is only due to interaction between the electronic spin $\bm{S}$ and the nuclear spin of Caesium $\bm{I}_{\mathrm{Cs}}$ (which form an intermediate angular moment $\bm{G_1}$) and that this interaction can be separated from the nuclear rotation. Without HFS interactions, the energy structure of the $e$ symmetry levels ($F_2$ levels in Hund's case (b) notation) is governed by the adiabatic PECs $U_X(R)$ and $U_a(R)$. The \astate\ state levels split into three $G_1$ HFS components and the splitting is given by diagonal matrix element of Fermi contact interaction term $A^{\mathrm{Cs}}_{1-1}$ for the $a^3\Sigma^+_1$ state:

\begin{equation}
A^{\mathrm{Cs}}_{1-1}(R)[G_1(G_1+1)-I(I+1)-S(S+1)]/2 \mbox{ ,}
\label{eq1}
\end{equation}

\noindent where $S=1$, $I=7/2$ while $G_1=5/2$, $7/2$ and $9/2$. The interaction between the $G_1=7/2$ component with the singlet \Xstate\ state levels is described by the off-diagonal matrix element:

\begin{equation}
A^{\mathrm{Cs}}_{0^+-1}(R)\sqrt{G_1(G_1+1)} \mbox{ .}
\label{eq2}
\end{equation}

Since the other two HFS components ($G_1=5/2$ and $G_1=9/2$) are not coupled to the \Xstate, effectively the \Xstate-- \astate\ HF interactions can be modeled by two channels as follows:

\begin{equation*}
\begin{blockarray}{ccc}
 &{^1\Sigma^+} &  {^3\Sigma^+_{G_1=7/2}} \\
\begin{block}{c(cc)}
 {^1\Sigma^+} \;\;           &U_X(R)\;\; &\frac{3\sqrt{7}}{2}A^{\mathrm{Cs}}_{0^+-1}(R)\\
 {^3\Sigma^+_{G_1=7/2}} \;\; &\frac{3\sqrt{7}}{2}A_{0^+-1}^{\mathrm{Cs}}(R)\;\;&U_a(R)-A^{\mathrm{Cs}}_{1-1}(R)\\
 \end{block}
\end{blockarray}
\end{equation*}

\noindent where the diagonal kinetic energy terms $-\hbar^2/(2\mu)(d^2/dR^2)$ and the rotational energy terms $\hbar^2J(J+1)/(2\mu R^2)$ are omitted for sake of clarity.

To construct the CC HFS Hamiltonian above we have used the empirical adiabatic potential energy curves for both singlet $U_X(R)$ and triplet $U_a(R)$ states as well as \emph{ab initio} HFS coupling functions $A^{\mathrm{Cs}}_{1-1}(R)$ and $A^{\mathrm{Cs}}_{0^+-1}(R)$ reported in Ref.~\cite{Krumins:2022}. The vibrational wave functions corresponding to $G_1=9/2$ and $G_1=5/2$ components of the \astate state could obtained under the conventional adiabatic approximation by means of the effective $U_a(R)+7/2A^{\mathrm{Cs}}_{1-1}(R)$ and $U_a(R)-9/2A^{\mathrm{Cs}}_{1-1}(R)$ interatomic PECs, respectively.


\begin{thebibliography}{10}

\bibitem{Carr:2009}
Krems R. V. Ye~J. Carr L.~D., DeMille~D.
\newblock {Cold and ultracold molecules: science, technology and applications}.
\newblock {\em New J Phys}, 11:055049, 2009.

\bibitem{Ulmanis:2012}
Ulmanis J., Deiglmayr J., Repp M., Wester R., and Weidem\"uller M.
\newblock {Ultracold Molecules Formed by Photoassociation: Heteronuclear
  Dimers, Inelastic Collisions, and Interactions with Ultrashort Laser Pulses}.
\newblock {\em Chem. Rev.}, 112:4890, 2012.

\bibitem{Quemener:2012}
P.~S.~Julienne G.~Qu\'em\'ener.
\newblock {Ultracold Molecules under Control!}
\newblock {\em Chem. Rev.}, 112:4949, 2012.

\bibitem{Ospelkaus:2008}
S.~Ospelkaus, A.~Pe$'$er, K.-K. Ni, J.~J. Zirbel, B.~Neyenhuis, S.~Kotochigova,
  P.~S. Julienne, J.~Ye, and D.~S. Jin.
\newblock {Efficient state transfer in an ultracold dense gas of heteronuclear
  molecules}.
\newblock {\em Nature Phys}, 4:622, 2008.

\bibitem{Field}
H\'el\`ene Lefebvre-Brion and Robert~W. Field.
\newblock {\em {The Spectra and Dynamics of Diatomic Molecules}}.
\newblock ELSEVIER Academic press, 2004.

\bibitem{Patel:2014}
H.~J. Patel, C.~L. Blackley, S.L. Cornish, and J.~M. Hutson.
\newblock Feshbach resonances, molecular bound states, and prospects of
  ultracold-molecule formation in mixtures of ultracold {K and Cs}.
\newblock {\em Phys. Rev. A}, 90(3):032716, 2014.

\bibitem{Borsalino:2016}
D.~Borsalino, R.~Vexiau, M.~Aymar, E.~{Luc-Koenig}, O.~Dulieu, and
  N.~{Bouloufa-Maafa}.
\newblock Prospects for the formation of ultracold polar ground state {KCs}
  molecules via an optical process.
\newblock {\em J. Phys. B: At. Mol. Opt. Phys.}, 49:055301, 2016.

\bibitem{Groebner:2017}
M.~Gr\"obner, Ph. Weinmann, E.~Kirilov, H.-Ch. N\"agerl, P.~S. Julienne,
  C.~R.~Le Sueur, and J.~M. Hutson.
\newblock Observation of interspecies {Feshbach} resonances in an ultracold
  {$^{39}$K-$^{133}$Cs} mixture and refinement of interaction potentials.
\newblock {\em Phys. Rev. A}, 95:022715, 2017.

\bibitem{Zukowski:2010}
J.~M.~Hutson, P.~S.~Zuchowski.
\newblock {Reactions of ultracold alkali-metal dimers}.
\newblock {\em Phys. Rev. A}, 81:060703, 2010.

\bibitem{Busevica:2011}
Busevica L., Klincare I., Nikolayeva O., Tamanis M., Ferber R., Meshkov~V. V.,
  Pazyuk~E. A., and Stolyarov~A. V.
\newblock Fourier transform spectroscopy and direct potential fit of a
  shelflike state: Application to {$E(4)^1\Sigma^+$} in {KCs}.
\newblock {\em J. Chem. Phys.}, 134:104307, 2011.

\bibitem{Szczepkowski:2012}
J.~Szczepkowski, A.~Grochola, W.~Jastrzebski, and P.~Kowalczyk.
\newblock On the {$4^1\Sigma^+$} state of the {KCs} molecule.
\newblock {\em J. Mol. Spectrosc.}, 276:19--21, 2012.

\bibitem{Klincare:2012}
Klincare I., Nikolayeva O., Tamanis M., Ferber R., Pazyuk E.A., and Stolyarov
  A.V.
\newblock Modeling of the {$X^1\Sigma^+$}, {$a^3\Sigma^+ \rightarrow
  E(4)^1\Sigma^+ \rightarrow  X^1\Sigma^+$ $(v=0,J=0)$} optical cycle for
  ultracold {KCs} molecule production.
\newblock {\em Phys. Rev. A}, 85(6):062520, 2012.

\bibitem{Ferber:2009}
R.~Ferber, I.~Klincare, O.~Nikolayeva, M.~Tamanis, H.~Kn\"ockel, E.~Tiemann,
  and A.~Pashov.
\newblock {$X^{1}\Sigma^{+}$} and {$a^{3}\Sigma^{+}$} states studied by
  {Fourier-transform} spectroscopy.
\newblock {\em Phys. Rev. A}, 80(6):062501, 2009.

\bibitem{Ferber:2013}
R.~Ferber, O.~Nikolayeva, M.~Tamanis, H.~Kn\"ockel, and E.~Tiemann.
\newblock Long-range coupling of {$X^{1}\Sigma^{+}$} and {$a^{3}\Sigma^{+}$}
  states of the atom pair {K} plus {Cs}.
\newblock {\em Phys. Rev. A}, 88(1):012516, 2013.

\bibitem{Krumins:2022}
V.~Krumins, M.~Tamanis, R.~Ferber, A.~V.~Oleynichenko, L.~V.~Skripnikov,
  A.~Zaitsevskii, E.~A.~Pazyuk, A.~V.~Stolyarov, and A.~Pashov.
\newblock The {$a^3\Sigma^+$} state of {KCs} revisited: hyperfine structure
  analysis and potential refinement.
\newblock {\em J. Quant. Spectrosc. Radiat. Transfer}, 283:108124, 2022.

\bibitem{Oleynichenko:2020}
A.~V. Oleynichenko, L.~V. Skripnikov, A.~Zaitsevskii, E.~Eliav, and V.~M.
  Shabaev.
\newblock Diagonal and off-diagonal hyperfine structure matrix elements in
  {KCs} within the relativistic {Fock} space coupled cluster theory.
\newblock {\em Chem. Phys. Lett.}, 756:137825, 2020.

\bibitem{Kato:1993}
{H. Kato}.
\newblock {Energy levels and Line intensities of Diatomic molecules.
  Application to Alkali Metal Molecules.}
\newblock {\em Bull. Chem. Soc. Jpn}, 66:3203, 1993.

\bibitem{Korek:2000}
M.~Korek, A.~R. Allouche, K.~Fakhreddine, and A.~Chaalan.
\newblock Theoretical study of the electronic structure of {LiCs, NaCs, and
  KCs} molecules.
\newblock {\em Can. J. Phys.}, 78(11):977--988, 2000.

\bibitem{Landau}
V.~I. Pupyshev, E.~A. Pazyuk, A.~V. Stolyarov, M.~Tamanis, and R.~Ferber.
\newblock {Analogue of oscillation theorem for nonadiabatic diatomic states:
  application to the A$^1\Sigma^+$ and b$^3\Pi$ states of KCs}.
\newblock {\em Physical Chemistry Chemical Physics}, 12(18):4809--4812, 2010.

\bibitem{FGH}
C.~C. Marston and G.~G. {Balint-Kurti}.
\newblock The {Fourier grid Hamiltonian} method for bound state eigenvalues and
  eigenfunctions.
\newblock {\em J. Chem. Phys.}, 91:3571--3576, 1989.

\bibitem{Havalyova:2021}
{I. Havalyova, I. Bozhinova, A. Pashov, A.J. Ross, P. Crozet}.
\newblock {A coupled-channels model describing the low-lying $^2\Delta$,
  $^2\Sigma^+$ and $^2\Pi$ electronic states of nickel monohydride with
  experimental accuracy}.
\newblock {\em J. Quant. Spetrosc. Radiat. Transfer}, 272:107800, 2021.

\bibitem{Warson2008}
James~K. G. Watson.
\newblock {H\"{o}nl-London factors for multiplet transitions in Hund's case
  \textbf{a} or \textbf{b}}.
\newblock {\em Journal of Molecular Spectroscopy}, 252(1):5--8, 2008.

\bibitem{Fraser}
N.E.Kuzmenko and A.~V. Stolyarov.
\newblock {Mathematical justification of the $R$-centroid approximation}.
\newblock {\em J. Quant. Spectrosc. Radiat. Transfer},
  35(5):415--418, 1986.

\bibitem{FCF_J}
A.~V. Stolyarov and N.E.Kuzmenko.
\newblock The influence of the rotation-vibrational interaction on the
  Franck-Condon factors for diatomic-molecules.
\newblock {\em Spectroscopy Letters}, 19(10):1113--1124, 1986.

\bibitem{Szczepkowski:2018}
J.~Szczepkowski, A.~Grochola, P.~Kowalczyk, and W.~Jastrzebski.
\newblock Spectroscopic study of the {(3)$C^1\Sigma^+ \leftarrow X^1\Sigma^+$}
  and {(2)$c^3\Sigma^+ \leftarrow X^1\Sigma^+$} transitions in {KCs} molecule.
\newblock {\em J. Quant. Spectrosc. Radiat. Transfer}, 204:131, 2018.

\bibitem{Kruzins:2021}
A.~Kruzins, V.~Krumins, M.~Tamanis, R.~Ferber, A.~V. Oleynichenko,
  A.~Zaitsevskii, E.~A. Pazyuk, and A.~V. Stolyarov.
\newblock Fourier-transform spectroscopy and relativistic electronic structure
  calculation on the {$c^3\Sigma^+$} state of {KCs}.
\newblock {\em J. Quant. Spectrosc. Radiat. Transf.}, 276:107902, 2021.

\bibitem{Manaa:2002}
M.~R. Manaa, A.~J. Ross, F. Martin, P.~Crozet, A.~M. Lyyra, L.~Li, C.~Amiot C.,
  and T.~Bergeman.
\newblock {Spin-orbit interactions, new spectral data, and deperturbation of
  the coupled {$b(1)^3\Pi_{\mathrm u}$} and {$A(1)^1\Sigma^+_{\mathrm u}$}
  states of {$K_2$}}.\newblock {\em J. Chem. Phys.}
\newblock 117:11208--11215, 2002.

\bibitem{Whiting_1974}
E.~E. Whiting and R.~W. Nicholls.
\newblock Reinvestigation of rotational-line intensity factors in diatomic
  spectra.
\newblock {\em Astrophys. J. Suppl.}, 27:1, 1974.

\bibitem{Brown_2003}
J. Brown and A. Carrington.
\newblock {\em Rotational Spectroscopy of Diatomic Molecules}.
\newblock Cambridge University Press, 2003.

\bibitem{Stolyarov:1994}
{A. V. Stolyarov and V. L. Popyshev}.
\newblock {Approximate sum rule for diatomic vibronic states}.
\newblock {\em Phys. Rev. A}, 49:1693, 1994.

\bibitem{Oleynichenko:2021}
A.~V. Oleynichenko and A.~Zaitsevskii.
Private communication, 2021.

\end{thebibliography}

\end{document}